\documentclass[12pt]{article}
\begin{document}
\title{Isospin fractionation in the nucleon emissions and fragment emissions 
in the intermediate energy heavy ion collisions }
\author{\small Jian-Ye Liu $^{1,2,3,5}$, Wen-Jun Guo $^{2}$, Yong Zhong Xing ,$^{1,3,5}$, 
Hang Liu  $^{4}$ \\}
\date{}
\maketitle $^{1}${\small Institute for the theory of modern physics, Tianshui Normal University, 
Gansu,Tianshui 741000, P. R. China}\\
$^{2}${\small Institute of Modern Physics, Chinese Academy of Sciences, P.O.Box 31,Lanzhou 730000,P.R.
China}\\
$^{3}${\small Center of Theoretical Nuclear Physics, National Laboratory of
Heavy Ion Accelerator Lanzhou 730000, P.R. China}\\
$^{4}${\small Department of Physics and Astronomy,Ohio University,Athens,OH45701,U.S.A}\\
$^{5}${\small CCAST(Word Lab.),P.O.Box 8730,Beijing 100080 Lanzhou 730000, P.R. China}\\

\date{}
\baselineskip 0.3in
\begin{center}{\bf Abstract}\end{center}
\hskip 0.3in  The degree of isospin fractionation is measured by $(N/Z)_{n}$ / $(N/Z)_{N_{imf}}$, 
where $(N/Z)_{n}$ and $(N/Z)_{N_{imf}}$ are the saturated neutron-proton ratio 
of nucleon emissions ( gas phase) and  that of fragment emissions (liquid phase)
in heavy ion collision at intermediate energy .  The calculated results by using the 
isospin-dependent quantum molecular dynamics model show that the degree of isospin 
fractionation is sensitive to the neutron-proton ratio of colliding system but
insensitive to the difference between the neutron-proton ratio of
target and that of projectile. In particular, the degree of isospin fractionation sensitively
 depends on the symmetry potential. However its dependences on the isospin dependent 
in-medium nucleon-nucleon cross section and momentum dependent interaction are rather weak.
 The nucleon emission (gas phase) mainly determines the dynamical 
behavior of the degree of isospin fractionation in the isospin fractionation process,
 compared to the effect of fragment emission. In this case, we propose that 
$(N/Z)_{n}$ / $(N/Z)_{N_{imf}}$ or $(N/Z)_{n}$  can be directly compared with the experimental
 data so that the information about symmetry potential can be obtained.\\\
{\bf PACS}: 25$\cdot$70$\cdot$pg,02$\cdot$70$\cdot$ $Ns$,
24$\cdot$10$\cdot$Lx\\\
\section{Indroduction}
The studies of the isospin effects at intermediate energy heavy ion collision can 
be used to get the information about isospin dependent in-medium nucleon-nucleon
cross section and isospin dependent mean field (symmetry potential). In order to 
obtain this information several interesting isospin effects in heavy ion collisions have
been explored both experimentally and theoretically over the last
few years${[1-16,26-28]}$.  Bao-An Li et al
investigated the isospin effect of the mean field and showed that
the neutron-proton ratio of preequilibrium nucleon emissions,the
neutron-proton differential flow and proton ellipse flow are
probes for extracting the information of the symmetry potential
${[1,2,14,15,16]}$. R.Pak et al have found that the isospin
dependences of collective flow and balance energy are mainly
originated from the isospin-dependent in-medium nucleon-nucleon
cross section ${[17,18]}$.  In order to extract above infoirmation 
 Bao-An Li ,M.Di Toro ,M. Colonna and V.Baran et al studied the isospin fractionation in the
intermediate energy heavy ion collisions in recent years
${[15,16,19,30]}$. The isospin fractionation is an unequal
partitioning of the neutron to proton ratio N/Z of unstable isospin
asymmetry nuclear matter in the low and high density regions.
Bao An Li 's work proposed a very useful point of review about isospin fractionation 
${[14]}$. V.Baran et al studied the isospin fractionation dynamics in details and 
gave more interesting results to explain  
the isopin fractionation dynamics${[30]}$. An indication of this phenomenon 
has been found recently in the intermediate energy heavy-ion experiments by H. Xu et al 
${[9]}$.\\\ 
 However the two essential ingredients in
heavy ion collision dynamics, the isospin dependent in-medium nucleon-nucleon
 cross section and the symmetry potential have not been well determined so far.
 In order to compare with experimental data directly it is important to quantify 
the gas phase and the liquid phase as well as their dependences on the dynamical ingredients 
in isospin fractionation process. Based on the 
isospin dependent quantum molecular dynamics model (IQMD)${[2,20,21,22,26]}$ we investigated
 the degree of isospin fractionation $(N/Z)_{n}$/$(N/Z)_{N_{imf}}$, where 
$(N/Z)_{n}$ and $(N/Z)_{N{imf}}$ are the saturated neutron to proton ratio of nucleon emissions 
and that of the fragment emissions respectively.  We investigated the dependeces of 
$(N/Z)_{n}$/$(N/Z)_{N{imf}}$ on the symmetry potential,
isospin dependent in-medium nucleon nucleon cross section and momentum dependent interaction  
 in some reaction processes. The calculated results show that $(N/Z)_{n}$/$(N/Z)_{N{imf}}$ only 
sensitively depends on symmetry potential for neutron-rich colliding system.
We simulated and discussed the dependences of isospin fractionation degree on the neutron 
proton ratio of colliding system and the difference between neutron proton ratio of projectile 
and that of target.  The roles of gas phase and liquid phase in the isospin fractionation process
are also investigated in detail.
\section{IQMD Model}
 As we have known that quantum molecular dynamics (QMD)${[21,22]}$ contains two dynamical 
ingredients: density dependent mean field and in-medium nucleon nucleon cross section. To 
describe isospin effects appropriately, QMD should be modified accordingly.  The density 
dependent mean field should contain correct isospin terms, including the symmetry potential 
and the Coulomb potential. The in-medium N-N cross section should identify neutron-neutron 
, proton-proton and neutron-proton collisions, in which Pauli blocking should be counted by 
distinguishing neutrons and protons.\\\ 
 Considering the above ingredients, we made important modifications
in QMD to obtain the isospin dependent quantum molecular dynamics (IQMD) 
. In IQMD model the initial density distributions of the colliding nuclei are from the
calculations by applying the Skyrme-Hatree-Fock model with the parameter set $SKM^{*}$$^{[23]}$. 
 The code of IQMD without the collision term is used to obtain the ground state properties of 
the colliding nuclei.  The ground state properties,such as binding energies and RMS radii etc, 
are consistent with the experimental data.  Thus the parameters of interaction potentials are 
fixed and used as input data in the dynamical calculations by  IQMD. In our
calculations the interaction potentials are taken as:
\begin{equation}
U(\rho)=U^{Sky}+U^{Coul}+U^{sym}+U^{Yuk}+U^{MDI}+U^{Pauli},
\end{equation}
where $U^{Sky}$, $U^{Coul}$, $U^{Yuk}$ and $U^{Pauli}$ are Skyrme
potential, Coulomb potential,Yukawa potential and Pauli
potential. $U^{MDI}$ is momentum dependent interaction with the form of
\begin{equation}
U^{MDI}=t_4ln^2[t_5(\overrightarrow{p_1}-\overrightarrow{p_2})^2+1]\frac
\rho {\rho _0}.
\end{equation} 
The more detailed physics ingredients and their numerical realization
in the IQMD model can be found in Refs 
[Chapter 10 in Ref.2,20,21,22,26]. 
$U^{sym}$ is the symmetry potential. In this paper, three different forms of $U^{sym}$
are used ${[1,15]}$,
\begin{equation}
U_1^{sym}=cF_1(u)\delta \tau _z
\end{equation}
\begin{equation}
U_2^{sym}=cF_2(u)[\delta \tau _z-\frac {1}{4}\delta ^2]
\end{equation}
\begin{equation}
U_3^{sym}=\frac{2cF_3(u)\delta\tau _z}{1+u}+\frac{cF_3(u)\delta^{2}}{(1+u)^{2}}
\end{equation}
 with
    \[\tau_{z}=\left\{ \begin{array}{ll}
              1 & \mbox{for neutron}\\
             -1 & \mbox{for proton}
             \end{array}
            \right. \]

Here c is the strength of symmetry potential, with the value of 32MeV,
 F$_1$(u)=u ,  $F_2(u)=u^{1/2}$ and F$_3$(u)=u$^2$, $u= \frac \rho {\rho_0}$.  $\delta $ 
is the relative neutron excess $\delta =\frac{\rho _n-\rho _p}{%
\rho _n+\rho _p}=\frac{\rho _n-\rho _p}\rho $.  $\rho $,$\rho _{_0}$,$%
\rho _n$ and $\rho _p$ are the total , normal , neutron
and proton densities respectively.\\\
 The curvatures $K_{sym}$ of $U_1^{sym}$ ,
$U_2^{sym}$ and $U_3^{sym}$  are -27 MeV, -69
MeV and 61 MeV respectively.\\\
 An empirical density dependent expression of the in-medium N-N cross section is
 ${[24]}$:
\begin{equation}
\sigma_{NN}=(1+\alpha\frac{\rho}{\rho_{0}})\sigma^{free}_{NN}
\end{equation}
with the parameter $\alpha$ $\approx-0.2$ which can well 
reproduce the flow data . Here $\sigma^{free}_{NN}$ is the
experimental nucleon nucleon cross section${[25]}$. The free
neutron-proton cross section is about 3 times larger
than the free proton-proton or the free neutron-neutron cross
section below about 400 MeV.  It should be pointed out that the ratio
$\sigma_{np}/\sigma_{pp}$ in the nuclear
medium sensitively depends on the evolutions of the nuclear density
distribution and beam energy. Here we made use of  equation (6) to
take the medium effect of two-body collision into account, in which 
the neutron-proton cross section is always larger than the
neutron-neutron or proton-proton cross section in the medium  at the beam energies
studied in this paper.\\\
\par
\section{Results and Discussions}
The isospin effect of the in-medium nucleon nucleon cross section on the
observables is defined by the difference between the observables from an
isospin dependent nucleon nucleon cross section
$\sigma^{iso}$ and those from an isospin independent nucleon
nucleon cross section $\sigma^{noiso}$.  The $\sigma^{iso}$ is
illustrated by $\sigma_{np} \geq \sigma_{nn}$=$\sigma_{pp}$ and
$\sigma^{noiso}$ means $\sigma_{np}$ = $\sigma_{nn}$ =
$\sigma_{pp}$  where $\sigma_{np}$, $\sigma_{nn}$, $\sigma_{pp}$ are 
neutron proton,neutron neutron and proton proton cross sections in medium respectively.
  In $(N/Z)_{n}/(N/Z)_{N_{imf}}$, the multiplicity of intermediate mass fragments $N_{imf}$ 
includes all fragments 
with the charge number from 2 to $(Z_{p}+Z_{t})/2$ , where $Z_{p}$ and $Z_{t}$ are 
the charge numbers of the projectile and target respectively .\\\
\subsection {The dependences of the degree of isospin fractionation on the symmetry 
potential and isospin dependent im-medium nucleon nucleon cross section}
Fig.1 shows the impact parameter averaged values of $<(N/Z)_{n}>_{b}/<(N/Z)_{N_{imf}>_{b}}$ 
as a function
 of the beam energy E in the reaction system  $^{124}Sn+^{124}Sn$. There are four different cases:
$U_1^{sym}$ + $\sigma^{iso}$ (solid line), $U_2^{sym}$ + $\sigma^{iso}$ (dashed line),
$U_2^{sym}$ + $\sigma^{noiso}$(dot-dashed line) and  $U_1^{sym}$ + $\sigma^{noiso}$ (dot line).

  It is found that for the different nucleon nucleon cross sections, $\sigma^{iso}$ and 
$\sigma^{noiso}$,  the variations among the ${(N/Z)_{n}}/(N/Z)_{imf}$
are smaller when the symmetry potential is fixed as either $U_2^{sym}$ or $U_1^{sym}$.
  However for different symmetry potentials $U_2^{sym}$ and $U_1^{sym}$ the gaps between the 
$(N/Z)_{n}/(N/Z)_{imf}$ are larger when the nucleon nucleon 
cross section is fixed as either $\sigma^{iso}$ or $\sigma^{noiso}$. So it might be concluded  
that ${(N/Z)_{n}}/(N/Z)_{imf}$ sensitively depends on the symmetry 
potential and weakly on the isospin effect of in-medium nucleon nucleon cross section. More 
complete understanding of the isospin fractionation processes can be obtained from the analysis
of the density dependent chemical potentials  for neutrons and protons ${[30]}$.  As one part
of chemical potential, the symmetry potential is repulsive for neutrons and  attractive for 
protons in the expanding process of the colliding system. This leads to the different chemical 
potential degradations with nuclear density $\rho_{q}$ for the neutron and proton in the low 
density region below normal nuclear density. However the mass flow in the nonequilibrium
expanding process is determined by the difference in the local values of the chemical 
potential and such difference directs the system from higher chemical potential region into 
lower values until equilibrium.  This transport effect induces the unstable isospin asymmeric
nuclear matter to be separated into a neutron-rich low density phase
and a neutron-poor high density phase for the neutron-rich system. Namely it is the different
chemical potential degradations with nuclear density for neutrons
and protons below normal nuclear density that induces the isospin fractionation.\\\ 
According to above analysis the symmetry potential in the chemical potential is a main 
dynamical ingredient for producing the isospin fractionation but the influence of two-body
collision on it is rather weak in the beam energy region in this paper.\\\
\subsection{The dependence of $(N/Z)_{n}/(N/Z)_{N_{imf}}$ on neutron to proton 
ratio of the colliding system}
 In order to investigate the evolution of $(N/Z)_{n}/(N/Z)_{N_{imf}}$ with the 
neutron to proton ratio of the colliding system , Fig.2 shows  
the $(N/Z)_{n}/(N/Z)_{N_{imf}}$ as a function of the neutron-proton ratio of the colliding
system for the reactions  $^{104}Sn+^{104}Sn$,$^{112}Sn+^{112}Sn$ and $^{124}Sn+^{124}Sn$ at
the beam energy E = 50 MeV/nucleon , impact parameter b=0.0 fm (right window) and 4.0 fm
(left window) .  The neutron to proton ratios of the above three colliding
systems are 1.01,1.34 and 1.48.  In Fig.2 there are three different
symmetry potentials $U_1^{sym}$ , $U_2^{sym}$  and $U_3^{sym}$. The calculation
results of the three reactions show that the $(N/Z)_{n}/(N/Z)_{N_{imf}}$ and its dependence
 on the symmetry potential are enhanced with increasing the neutron to proton ratio of the
colliding systems. In particular, the $(N/Z)_{n}/(N/Z)_{N_{imf}}$ is larger than 1.0 for the 
neutron-rich system 
$^{124}Sn+^{124}Sn$ ,which means the neutron-rich gas phase and the neutron-poor liquid phase 
,compared to the neutron to proton ratio of colliding system. It is also found that  
the $(N/Z)_{n}/(N/Z)_{N_{imf}}$ is enhanced with increasing
symmetry potential strength. The amplitudes of $(N/Z)_{n}/(N/Z)_{N_{imf}}$ from large to 
small correspond to the order of symmetry potential strength. The order of symmetry
potential strengthes in the lower density region below the normal nuclear density 
are from $U_2^{sym}$ (dot line) , $U_1^{sym}$ (solid line) to $U_3^{sym}$ (dot-dashed line)
 for the neutron-rich system. The order of symmetry potential strength below the normal
 density is contrary to those at the normal nuclear density (see Fig.4.1 in ${[1]}$).\\\ 
However for the neutron-poor systems, such as, $^{104}Sn+^{104}Sn$ and $^{112}Sn+^{112}Sn$ 
the values of $(N/Z)_{n}/(N/Z)_{N_{imf}}$ are  decreased gradually below 1.0 with decreasing the 
neutron-proton ratio of colliding system. This means the neutron-poor gas phase and 
neutron-rich liquid phase, compared to the neutron proton ratio of colliding system. 
In the isospin fractionation process for the neutron-poor system, 
$(N/Z)_{n}$ is decearsed due to
 neutron-poor colliding system. On the other hand , $(N/Z)_{imf}$ 
is increased, compared to the initial neutron proton ratio of neutron-poor 
colliding system.  This can be explained by the fact that the excited primary neutron-poor 
fragments decays to a stable neutron-rich one which is  closer to the beta stable line.   
\subsection {The dependences of $<(N/Z)_{n}>_{b}/<(N/Z)_{N_{imf}}>_{b}$ on the gas phase and 
liquid phase}
In order to study the dependences of $<(N/Z)_{n}>_{b}/<(N/Z)_{N_{imf}}>_{b}$ on the gas phase 
and liquid phase, Fig.3 shows the impact parameter averaged values of $<(N/Z)_{n}>_{b}$ (in the 
left panel) and $<(N/Z)_{N_{imf}}>_{b}$ (in the right panel) as a function of the beam energy E
 for the different symmetry potentials $U_1^{sym}$ and $U_2^{sym}$. The incident channel 
conditions and the line symbols in Fig.3 are the same as those in Fig.1.  It is clear to see that the 
dynamical behavior of $<(N/Z)_{n}>_{b}$ (gas phase) is very similar to that of 
 $<(N/Z)_{n}>_{b}/<(N/Z)_{N_{imf}}>_{b}$ in Fig.1. However all lines  for 
$<(N/Z)_{N_{imf}}>_{b}$ (liquid phase) corresponding the different symmetry potentials 
$U_1^{sym}$ and $U_3^{sym}$ as well as different nucleon nucleon cross sections $\sigma^{iso}$ 
and $\sigma^{noiso}$ are very close to each other. Namely $<(N/Z)_{N_{imf}}>_{b}$ is insensitive
 to both the symmetry potential and the isospin effect of in-medium nucleon nucleon cross section.  
So the nucleon emissions (gas phase) mainly  determine the dynamical behavior of 
the degree of isospin fractionation for the neutron-rich colliding system in the beam energy 
region studied here.\\\
It should be pointed out that the calculated results in above three figures are taken by 
 accounting the time up to 300 fm/c. This time roughly corresponds to the freeze-out time because
 during fragmentation stage the colliding system breaks up and the fragment formation
 process occurs up to the time around 240-300 fm/c in 
the beam energy region studied here. From the time evolusion of impact parameter averaged value
 of nucleon emission number $<N_{n}>_{b}$ and that of fragment emission number
$<N_{imf}>_{b}$ in the reaction $^{124}Sn+^{124}Sn$ at 
E= 50 MeV/nucleon shown in Fig.4.b, we can see that $<N_{imf}>_{b}$ 
 almost stabilizes after 250 fm/c, even though $<N_{n}>_{b}$  continues to be enhanced very 
slowly after 250 fm/c. However the impact parameter averaged value of the neutron proton ratio
 of nucleon emission and that of fragment emission $<(N/Z)_{n}>_{b}$ and  
$<(N/Z)_{N_{imf}}>_{b}$ shown in Fig.4.a are almost constant after 200 fm/c.\\\  
From above analysis and discussion in Fig.4 we might 
conclude that even though the simulation by IQMD could not 
include all secondary decaies, it does not matter too much on the values 
 of $<(N/Z)_{n}>_{b}$ and $<(N/Z)_{N_{imf}}>_{b}$ at 300 fm/c and the conclusions.
  Furthermore the observable for comparing with the experimental data in this paper is 
$<(N/Z)_{n}>_{b}$/$<(N/Z)_{N_{imf}}>_{b}$  .
\subsection{The dependence of $(N/Z)_{n}/(N/Z)_{N_{imf}}$ on the difference between
neutron to proton ratios for projectile and target}
Fig.5 shows the time evolution of $(N/Z)_{n}/(N/Z)_{N_{imf}}$ in the reaction systems which 
have different neutron-proton
ratios for target and projectile but with the same mass and  the
same neutron-proton ratio of colliding system ,the same $U_1^{sym}$ 
and the same $\sigma^{iso}$ at beam energy E=50 MeV/nucleon and impact parameter of 2.0 fm
. In Fig.5 there are two couples of reactions $^{74}Zn+^{74}Se$ and 
$^{74}Ge+^{74}Ge$ ( left window) as well as $^{132}Sn+^{112}Sn$ and $^{124}Sn+^{124}Sn$ 
(right window). The neutron-proton ratios of projectile, target and colliding system
are listed in table 1.  
\begin{center}
\scriptsize
\begin{tabular}{|c|c|c|c|c|c|c|} \hline
\small
 Nucleus&$^{74}Zn$&$^{74}Se$&$^{74}Ge$&$^{116}Sn$&
$^{132}Sn$&$^{124}Sn$\\\hline
 N/Z&1.47&1.18&1.31&1.32&1.64&1.48 \\\hline
\end{tabular}\\
\end{center}
\begin{center}
\scriptsize
\begin{tabular}{|c|c|c|c|c|} \hline
\small
  system & $^{74}Zn+^{74}Se$ & $^{74}Ge+^{74}Ge$ & $^{132}Sn+^{116}Sn$ 
& $^{124}Sn+^{124}Sn$\\\hline
  N/Z & 1.31 & 1.31 & 1.48 & 1.48 \\\hline
\end{tabular}\\
\end{center}
From Fig.5 it is clear to see that all solid lines and all 
dashed lines are very close to each other for the same mass and 
the same neutron proton ratio of each couple of colliding systems. It shows that the
difference between the neutron-proton ratios for target and
projectile doesn't influence isospin fractionation process,
instead it sensitively depends on the neutron-proton ratio of colliding dsystem
as observed in Fig.2 \\\
It is worth mentioning that because the scaling time of the
isospin fractionation process is faster than that of fragmentation
process , $(N/Z)_{n}$ is much larger than $(N/Z)_{N_{imf}}$ before 125 fm/c
. This indicates that the large fluctuation before 100 fm/c, then $(N/Z)_{n}$ and 
$(N/Z)_{N_{imf}}$ gradually approach their equilibrium values with the
increasing colliding time. This is why the beginning time of the time evolutions for figures
is started at 100 fm/c.\\\
\subsection{The role of MDI on the dynamical process of isospin fractionation }
 The nonlocal property of the nuclear interaction leads to a
repulsive momentum dependent interaction ($U^{MDI}$ ) in the
intermediate energy heavy ion collisions. In order to study the influence
of the $U^{MDI}$ on the dynamical process of isospin fractionation Fig.6 shows 
the time evolution of impact parameter averaged values of 
$<(N/Z)_{n}>_{b}/<(N/Z)_{N_{imf}}>_{b}$ with the same
$\sigma^{iso}$ but different symmetry potentials $U_1^{sym}$ and $U_2^{sym}$.
In calculations we consider momentum dependent interaction $U^{MDI}$ and momentum independent
interaction $U^{noMDI}$ for the reaction $^{124}Sn+^{124}Sn$ at E= 50 MeV/nucleon. There are
four cases: $U_1^{sym}$ +$U^{MDI}$ ; $U_1^{sym}$ +$U^{noMDI}$ 
;$U_2^{sym}$ +$U^{MDI}$  and $U_2^{sym}$ +$U^{noMDI}$ . The gap between the values of
$<(N/Z)_{n}>_{b}/<(N/Z)_{N_{imf}}>_{b}$ is smaller for the $U^{MDI}$  and $U^{noMDI}$ with
the same $U_1^{sym}$ or $U_2^{sym}$.  However,the gap
between the values of $<(N/Z)_{n}>_{b}/<(N/Z)_{N_{imf}}>_{b}$  is lager for the same
the $U^{MDI}$ or $U^{noMDI}$  but different symmetry potentials. It turns out that the influence
 of $U^{MDI}$  on the isospin
fractionation process is not obvious because the emission
probabilities induced by  $U^{MDI}$ are about the same for the neutrons and protons .\\\
\section{Summary and conclusions}
 In summary,from the calculation results we can get following conclusions:
(1)The $<(N/Z)_{n}>_{b}/<(N/Z)_{N_{imf}}>_{b}$ is enhanced with
increasing the neutron to proton ratio of colliding system and symmetry potential strength 
 for neutron-rich system. However, the isospin fractionation process is insensitive to all these
ingredients for the neutron-poor system in beam energy region studied here.\\\  
(2) In particular,$<(N/Z)_{n}>_{b}/<(N/Z)_{N_{imf}}>_{b}$  and  $<(N/Z)_{n}>_{b}$ sensitively
depend on the symmetry potential but weakly on the isospin effect of in-medium 
nucleon-nucleon cross section and momentum dependent interaction for the neutron-rich system.\\\
(3)The dynamical behavior of $<(N/Z)_{n}>_{b}/<(N/Z)_{N_{imf}}>_{b}$ is mainly determined by 
the nucleon emissions (gas phase). \\\ 
(4) The $(N/Z)_{n}/(N/Z)_{N_{imf}}$ sensitively 
depends on the neutron to proton ratio of colliding system but weakly on the
difference between the neutron proton ratios for target and projectile.\\\ 
\section{ACKNOWLEDGMENT}

We thank Prof.Bao-An Li,Che Ming Ko and Dr.Lie Wen Chen for helpful discussions.\\\
This work is supported by the Major State Basic Research Development Program
in China Under Contract No.G2000077400, "100-person project" of the Chinese
Academy of Sciences, the National Natural Science Foundation of China under Grants
Nos.10175080,10175082, 10004012,19847002 and The CAS Knowledge Innovation Project
 No.KJCX2-SW-N02


\baselineskip 0.2in
\section*{Figure captions}
\begin{description}
\item [Fig.1]
The impact parameter averaged value of $<(N/Z)_{n}>_{b}/<(N/Z)_{N_{imf}>_{b}}$ as a function
of the beam energy E for the reaction system $^{124}Sn+^{124}Sn$ and the different symmetry
potentials $U_1^{sym}$ and $U_2^{sym}$ with $\sigma^{iso}$ or $\sigma^{noiso}$.
\item[Fig.2]
The $(N/Z)_{n}/(N/Z)_{N_{imf}}$ as a function of the
neutron-proton ratio of colliding systems
$^{104}Sn+^{104}Sn$,$^{112}Sn+^{112}Sn$ and $^{124}Sn+^{124}Sn$
for three different symmetry potentials  $U_1^{sym}$ ,$U_2^{sym}$
and  $U_3^{sym}$ at the beam energy of 50 MeV/nucleon and 
impact parameter of 4.0 fm (left panel) and 0.0 fm (right panel).
\item[ Fig.3]  The impact parameter averaged
values of $<(N/Z)_{n}>_{b}$ (left panel) and $<(N/Z)_{N_{imf}}>_{b}$  
(right panel ) as a function of beam energy  E for the different symmetry potentials 
$U_1^{sym}$ and $U_2^{sym}$ as well as the reaction $^{124}Sn+^{124}Sn$.
\item[ Fig.4] 
The time evolusions of impact parameter averaged values of nucleon emission number 
$<N_{n}>_{b}$ and fragment emission number $<N_{imf}>_{b}$ 
(Fig.4.b)) as well as that of neutron proton ratio of nucleon emission  $<(N/Z)_{n}>_{b}$ and 
that of fragment emission $<(N/Z)_{N_{imf}}>_{b}$ (Fig.4.a) for the reaction $^{124}Sn+^{124}Sn$
 at E= 50 MeV/nucleon. 
\item[Fig. 5]
Time evolution of $(N/Z)_{n}/(N/Z)_{N_{imf}}$ for different
neutron-proton ratios for target and projectile but the same mass
and the same neutron proton ratio of each couple of colliding systems for
the reactions $^{74}Zn+^{74}Se$ and $^{74}Ge+^{74}Ge$ (left
window) as well as  $^{132}Sn+^{112}Sn$ and $^{124}Sn+^{124}Sn$ (right
window).
\item[ Fig.6]
the time evolutions of the impact parameter averaged values of
$<(N/Z)_{n}>_{b}/<(N/Z)_{N_{imf}}>_{b}$ with the same $\sigma^{iso}$ ,the same  $U^{MDI}$ 
or $U^{noMDI}$ but different symmetry potentials  $U_1^{sym}$ and $U_2^{sym}$ 
for the reaction $^{124}Sn+^{124}Sn$ at the beam energy of 50 MeV/nucleon.
\end{description}

\end{document}